# AdaMSS: Adaptive Multi-Modality Segmentation-to-Survival Learning for Survival Outcome Prediction from PET/CT Images

Mingyuan Meng [a, b, ‡], Bingxin Gu [c, ‡], Michael Fulham [a, d], Shaoli Song [c, *], Dagan Feng [a], Lei Bi [b, *], and Jinman Kim [a, *]

[a] School of Computer Science, The University of Sydney, Australia.
[b] Institute of Translational Medicine, Shanghai Jiao Tong University, Shanghai, China.
[c] Department of Nuclear Medicine, Fudan University Shanghai Cancer Center, Shanghai, China; Department of Oncology, Shanghai Medical College, Fudan University, Shanghai, China; Center for Biomedical Imaging, Fudan University, Shanghai, China; Shanghai Engineering Research Center of Molecular Imaging Probes, Shanghai, China; Key Laboratory of Nuclear Physics and Ion-Beam Application (MOE), Fudan University, Shanghai, China.
[d] Department of Molecular Imaging, Royal Prince Alfred Hospital, Australia.

**Abstract** — Survival prediction is a major concern for cancer management. Deep survival models based on deep learning have been widely adopted to perform end-to-end survival prediction from medical images. Recent deep survival models achieved promising performance by jointly performing tumor segmentation with survival prediction, where the models were guided to extract tumor-related information through Multi-Task Learning (MTL). However, these deep survival models have difficulties in exploring out-of-tumor prognostic information (e.g., lymph node metastasis and heterogeneity in tumor periphery). In addition, existing deep survival models are unable to effectively leverage multi-modality images (e.g., PET/CT). Empirically-designed fusion strategies (e.g., early/late fusion) were commonly adopted to fuse multi-modality information via task-specific manually-designed networks, thus limiting the adaptability to different scenarios. In this study, we propose an Adaptive Multi-modality Segmentation-to-Survival model (AdaMSS) for survival prediction from PET/CT images. Instead of adopting MTL, we propose a novel Segmentation-to-Survival Learning (SSL) strategy, where our AdaMSS is trained for tumor segmentation and survival prediction sequentially in two stages. This strategy enables the AdaMSS to focus on tumor regions in the first stage and gradually expand its focus to include other prognosis-related regions in the second stage. We also propose a data-driven strategy to fuse multi-modality information, which realizes adaptive optimization of fusion strategies based on training data during training. With the SSL and data-driven fusion strategies, our AdaMSS is designed as an adaptive model that can self-adapt its focus regions and fusion strategy for different training stages. Our AdaMSS is also capable of incorporating conventional radiomics features as an enhancement, where handcrafted features can be extracted from the AdaMSS-segmented tumor regions and then integrated into the AdaMSS through cooperative training and inference. Extensive experiments with two large clinical datasets (including 1,380 patients acquired from nine medical centers) show that our AdaMSS outperforms state-of-the-art survival prediction methods.

## 1. Introduction

Survival prediction is crucial for cancer management as it can provide early prognostic information for personalized treatment planning (Deepa and Gunavathi, 2022). Survival prediction is a regression task that aims to predict the survival outcomes of patients such as Progression-Free Survival (PFS) and Recurrence-Free Survival (RFS). As patients may be lost during follow-up, survival prediction often encounters right-censored survival data: it is only known that the patients survived for a period of time but the exact time of event occurrence (e.g., disease progression) is unclear, which incurs difficulties in designing survival models to leverage this incomplete data. In addition, the survival outcomes of patients are influenced by complex interactions among a variety of factors including treatment regimens, clinical demographics, and the underlying disease physiology (Naser et al., 2022). This makes survival prediction an intractable challenge that tends to integrate multi-modality data (e.g., PET/CT images) to optimize prediction performance (Deepa and Gunavathi, 2022; Naser et al., 2022). In recent years, survival prediction has gained wide attention (Deepa and Gunavathi, 2022; Doja et al., 2020), especially for cancers with high incidence or mortality such as Head and Neck (H&N) cancers (Sung et al., 2021). For example, the HEad and neCK TumOR segmentation and outcome prediction (HECKTOR) challenge has been held for three years to facilitate the development of survival outcome prediction methods for H&N cancers from PET/CT images (Andrearczyk et al., 2022; Andrearczyk et al., 2023).

Traditional survival prediction methods are usually based on radiomics (Gillies et al., 2015), in which high-dimensional handcrafted features are extracted from medical images and modeled by statistical survival models such as Cox Proportional Hazard (CoxPH) model (Cox et al., 1972). Radiomics has been widely used in survival prediction for its capability to precisely characterize intra-tumor information including tumor texture, intensity, heterogeneity, and morphology (Lambin et al., 2012; Sollini et al., 2019). However, as radiomics features are extracted from segmented regions that are often limited to primary tumors (Zhang et al., 2017; Lv et al., 2019), radiomics has difficulties in characterizing the prognostic information outside primary tumors, such as lymph node metastasis (Kawada and Taketo, 2011) and heterogeneity in tumor periphery (Erber et al., 2022). There have been studies that incorporate lymph node segmentation in radiomics analysis (Lu et al., 2021; Salahuddin et al., 2023); however, lymph node segmentation is a challenging task and tumor peripheral heterogeneity was rarely investigated. Deep survival models based on deep learning have shown the potential to address this limitation as they can perform end-to-end prediction from medical images without necessarily requiring tumor masks (Deepa and Gunavathi, 2022), thus potentially leveraging the prognostic information throughout the entire image for survival prediction. Unfortunately, performing end-to-end prediction without tumor masks brings in interference from non-relevant backgrounds and incurs difficulties in extracting tumor-specific information. There also exist deep survival models that take tumor masks as input to exclude non-tumor regions (Zhang et al., 2020; Gu et al., 2022), but these models have lost the potential to leverage out-of-tumor prognostic information.

The above dilemma exposes a crucial question for survival prediction: how to capture as much prognostic information as possible while getting less interference from non-relevant information. Recently, deep survival models attempted to answer this question by Multi-Task Learning (MTL) (Zhao et al., 2022) and have demonstrated state-of-the-art prediction performance (Andrearczyk et al., 2021; Meng et al., 2022a; Saeed et al., 2022a; Wu et al., 2023). Tumor segmentation has been shown as an effective auxiliary task for survival outcome prediction in various cancers including Oropharynx Cancer (OPC) (Andrearczyk et al., 2021), Nasopharyngeal Carcinoma (NPC) (Meng et al., 2022a), and brain gliomas (Wu et al., 2023). Through jointly learning tumor segmentation, deep survival models can be implicitly guided to extract tumor-related information while not absolutely excluding out-of-tumor information in images. However, these models cannot freely leverage the out-of-tumor prognostic information due to the destructive interference (Yu et al., 2020), also known as negative transfer (Liu et al., 2019), intrinsic to MTL: survival prediction expects models to encompass out-of-tumor prognostic information, while tumor segmentation expects models to focus on tumor regions.

How to leverage multi-modality medical images is another crucial question for survival prediction as multi-modality images can provide potentially complementary information (Zhou et al., 2020; D'Souza et al., 2022). For example, PET/CT images can provide both anatomical (from CT) and glucose metabolic (from PET) information about tumors, which have been shown to benefit survival prediction (Gu et al., 2022). However, existing deep survival models are sub-optimal in utilizing multi-modality images. Empirically-designed fusion strategies (e.g., early/late fusion) were commonly adopted to fuse multi-modality information using task-specific manually-designed networks (Zhou et al., 2020; D'Souza et al., 2022; Meng et al., 2022a; Saeed et al., 2022a). The empirically-designed fusion strategies, however, are inherently limited in adaptability and tend to be sub-optimal, because the possible search space for fusion strategies can be too large to be manually searched and the optimal strategy also could vary depending on the data (e.g., different diseases) and the tasks (e.g., segmentation or survival prediction).

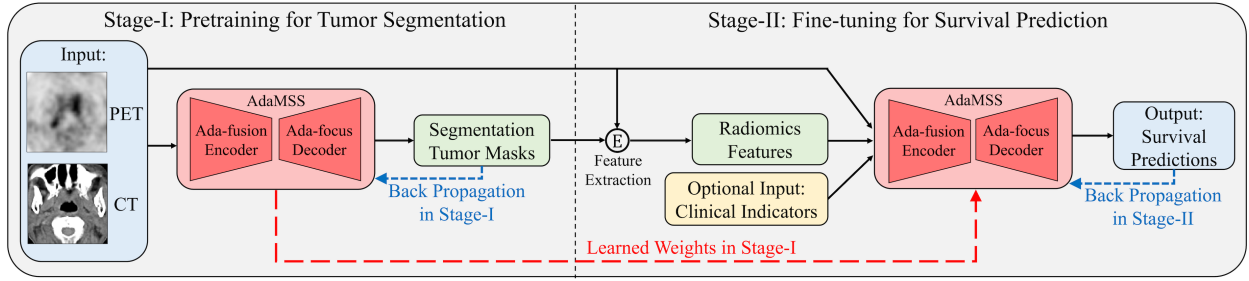

Fig. 1. The workflow of our Adaptive Multi-modality Segmentation-to-Survival model (AdaMSS), which takes PET/CT images as input and predicts the survival outcomes of patients. Clinical indicators, if available, are optional input to further improve survival prediction. The AdaMSS is first pretrained for tumor segmentation (Stage-I) and then fine-tuned for survival prediction (Stage-II). The tumor masks produced in Stage-I can be leveraged to extract handcrafted radiomics features, so as to facilitate the survival prediction fine-tuning in Stage-II.

To cope with the two crucial questions for survival prediction, we propose an Adaptive Multi-modality Segmentation-to-Survival model (AdaMSS) for survival prediction from PET/CT images. The technical contributions in the AdaMSS are three folds: (i) We propose a novel Segmentation-to-Survival Learning (SSL) strategy, where the AdaMSS is first pretrained for tumor segmentation and then fine-tuned for survival prediction in two training stages (Fig. 1). This learning strategy is free from the destructive interference intrinsic to MTL and enables our AdaMSS to initially focus on tumor regions and then gradually expand its focus to include other prognosis-related regions. Moreover, the SSL strategy also provides flexibility to incorporate conventional radiomics features as an enhancement for survival prediction, where handcrafted image features can be extracted from the AdaMSS-segmented tumor regions and then integrated into the AdaMSS through cooperative training and inference (Fig. 1). (ii) We propose a data-driven strategy to fuse multi-modality information, which realizes adaptive optimization of fusion strategies based on training data during training. To realize this, multi-modality information is allowed to be fused at many potential network locations and then a fusion gate module is proposed to control how to fuse information at each location, which makes a fundamental shift from existing empirically-designed strategies to a data-driven strategy. (iii) With the proposed SSL and data-driven fusion strategies, we design the AdaMSS as an adaptive model that can self-adapt its focus regions and fusion strategy for different training targets (i.e., tumor segmentation or survival prediction) in different training stages. Extensive experiments with two large clinical datasets (including 1,380 OPC/NPC patients acquired from nine medical centers) demonstrate that our AdaMSS outperforms state-of-the-art survival prediction methods, including the top-performing methods in the HECKTOR 2021/2022 challenge.

## 2. Related Work
### 2.1. Radiomics-based Survival Prediction

The CoxPH model (Cox et al., 1972) is the most widely used statistical survival model and has been extended to many variants, such as the Lasso-CoxPH (Tibshirani, 1998) that uses Least Absolute Shrinkage and Selection Operator (LASSO) regression for feature reduction. Recently, an Individual Coefficient Approximation for Risk Estimation (ICARE) model (Rebaud et al., 2023) was proposed and achieved one of the top performance in the HECKTOR 2022 challenge. As these statistical survival models cannot directly apply to image data, handcrafted radiomics features are necessary for traditional survival prediction methods.

Zhang et al. (2017) explored survival prediction for NPC from MRI images, where radiomics features were extracted from manually-delineated primary tumor regions and then modeled by a Lasso-CoxPH model. Lv et al. (2019) also explored the same problem with PET/CT images, where the extracted radiomics features were modeled by a CoxPH model. For OPC, Rebaud et al. (2023) extracted radiomics features from PET/CT images and built an ICARE model. As previously mentioned, these traditional methods relied on handcrafted radiomics features that were limited to the segmented tumor regions.

## 2.2. Deep Learning for Survival Prediction

Early deep survival models focus on learning the complex relevance between clinical prognostic indicators and survival outcomes (Katzman et al., 2018; Gensheimer and Narasimhan, 2019), which cannot directly apply to image data and require radiomics features to characterize images. Then, end-to-end deep survival models were developed to perform survival prediction directly from medical images (Deepa and Gunavathi, 2022). For example, Zhang et al. (2020) proposed an end-to-end deep survival model for survival prediction from CT images. This model takes manually segmented tumor regions as input and disregards the prognostic information outside primary tumors. Jing et al. (2020) proposed a Multi-modality Deep Survival Network (MDSN) for survival prediction in NPC from MRI images. This model takes images covering the whole nasopharynx as input, which enables it to access complete image information but brings in interference from non-relevant backgrounds.

Recently, multi-task deep survival models were proposed and achieved state-of-the-art performance (Andrearczyk et al., 2021; Meng et al., 2022a; Saeed et al., 2022a; Wu et al., 2023). Andrearczyk et al. (2021) proposed to use a hard-sharing Unet for joint tumor segmentation and survival prediction. Saeed et al. (2022a) proposed a Transformer-based Multimodal network for Segmentation and Survival prediction (TMSS) using a hard-sharing Visual Transformer (ViT) (Dosovitskiy et al., 2021). In our previous studies (Meng et al., 2022a; Meng et al., 2022b), we proposed a Deep Multi-Task Survival model (DeepMTS) that uses a hybrid multi-task framework for joint tumor segmentation and survival prediction. The benefits of jointly learning tumor segmentation and survival prediction have been well demonstrated, but few studies explored the intrinsic destructive interference between them. From our review, this is the first study that identifies this destructive interference and hence introduces the SSL strategy as an improvement.

Deep survival models can also be leveraged along with conventional radiomics features and/or statistical survival models. Saeed et al. (2022b) proposed to combine a Deep Multi-Task Logistic Regression model with a CoxPH model (DeepMTLR-CoxPH), which achieved top performance in the HECKTOR 2021 challenge. We previously attempted to enhance DeepMTS with radiomics features and developed a radiomics-enhanced DeepMTS (Radio-DeepMTS) (Meng et al., 2023), which was placed 2[nd] in the HECKTOR 2022 challenge with a negligible performance difference from the 1[st] place (0.00068 in C-index with $P > 0.05$). Note that these methods integrated statistical survival models and deep survival models at the decision level and did not consider the synergies during training, while our SSL strategy provides feasibility to incorporate radiomics features into AdaMSS for cooperative training.

In addition, existing deep survival models usually were developed and evaluated for a single disease (Jing et al., 2020; Zhou et al., 2020; Andrearczyk et al., 2021; Cui et al., 2022; D'Souza et al., 2022; Meng et al., 2022a; Saeed et al., 2022a; Tang et al., 2022; Wu et al., 2023), which might incur suspicion on their adaptability and generalizability to other diseases. To alleviate this suspicion, our AdaMSS was evaluated for two diseases (OPC and NPC) with two large multi-center clinical datasets.

## 2.2. Survival Prediction with Multi-modality Images

Multi-modality images have been widely used for survival prediction. For radiomics-based survival prediction methods, radiomics features can be separately extracted from different imaging modalities and then fed into statistical survival models for integration (Lv et al., 2019; Rebaud et al., 2023). For deep survival models, how to combine multi-modality information could be more complicated. Early and late fusion are the most common strategies, where multi-modality images are concatenated as multi-channel inputs (early fusion) (Jing et al., 2020; Andrearczyk et al., 2021; Meng et al., 2022a; Naser et al., 2022; Saeed et al., 2022a; Meng et al., 2023) or enter multiple independent encoders with resultant features fused (late fusion) (Zhou et al., 2020; Cui et al., 2022; D'Souza et al., 2022). However, early fusion has difficulties in extracting intra-modality information due to concatenated images for feature extraction, while late fusion has difficulties in extracting inter-modality information due to fully independent feature extraction. Recently, Tang et al. (2022) proposed an intermediate fusion strategy for survival prediction, where multi-modality images were fed into multiple encoders with intermediate features fused at multiple scales via sophisticated fusion modules.

Intermediate fusion strategies have also been leveraged in other medical applications, such as skin lesion classification (Bi et al., 2020) and ventricle segmentation (Sun et al., 2022), and showed better utilization/integration of multi-modality images. Nevertheless, as mentioned above, these fusion strategies were empirically designed and inherently limited. Based on our review, this is the first study that proposes a data-driven fusion strategy for survival prediction.

## 3. Method

### 3.1. Overview

Our AdaMSS is designed as an adaptive model for both tumor segmentation and survival prediction. Fig. 2 presents the architecture of AdaMSS, consisting of an adaptive fusion (Ada-fusion) encoder and an adaptive focus (Ada-focus) decoder. The Ada-fusion encoder extracts features from PET/CT images using our data-driven fusion strategy (detailed in Section 3.2). The Ada-focus decoder adopts attention gate modules (Schlemper et al., 2019) to adaptively focus on task-related target regions (detailed in Section 3.3). The AdaMSS is trained by our SSL strategy (detailed in Section 3.4), where it is first pretrained for tumor segmentation (Stage-I) and then fine-tuned for survival prediction (Stage-II). The AdaMSS is adaptive to different training targets in different training stages because it can adapt its fusion strategy and focus regions during training. In addition, the segmentation masks produced in Stage-I can be leveraged to extract radiomics features so as to facilitate survival prediction in Stage-II.

### 3.2. Adaptive Fusion (Ada-fusion) Encoder

The Ada-fusion encoder includes three branches for feature extraction and fusion. Two branches, named $B_{PET}$ and $B_{CT}$, extract features from PET and CT images separately, while the other branch, named $B_{Fuse}$, first extracts features from concatenated PET/CT images and then fuse the features from $B_{PET}$ and $B_{CT}$ using fusion gate modules. Each branch contains successive residual blocks with max-pooling layers applied between adjacent blocks to reduce feature map resolution. Each residual block consists of $n$ stacked $3 \times 3 \times 3$ convolutional layers with residual connections and followed by Batch Normalization (BN) and ReLU activation. Fusion gate modules are embedded after each residual block of $B_{Fuse}$ for feature fusion. Formally, let $F_{pet}^i$, $F_{ct}^i$, and $F_{fuse}^i$ be the features from the $i^{th}$ residual block of $B_{PET}$, $B_{CT}$, and $B_{Fuse}$. Let $\mathcal{F}_i$ be the fusion gate module after the $i^{th}$ residual block and $F_{\mathcal{F}}^i$ be the fused features from the $\mathcal{F}_i$. We can derive $F_{\mathcal{F}}^i = \mathcal{F}_i(F_{pet}^i, F_{ct}^i, F_{fuse}^i)$, which is fed into the next residual block of $B_{Fuse}$ for later fusion and also is propagated to the Ada-focus decoder through skip connections.

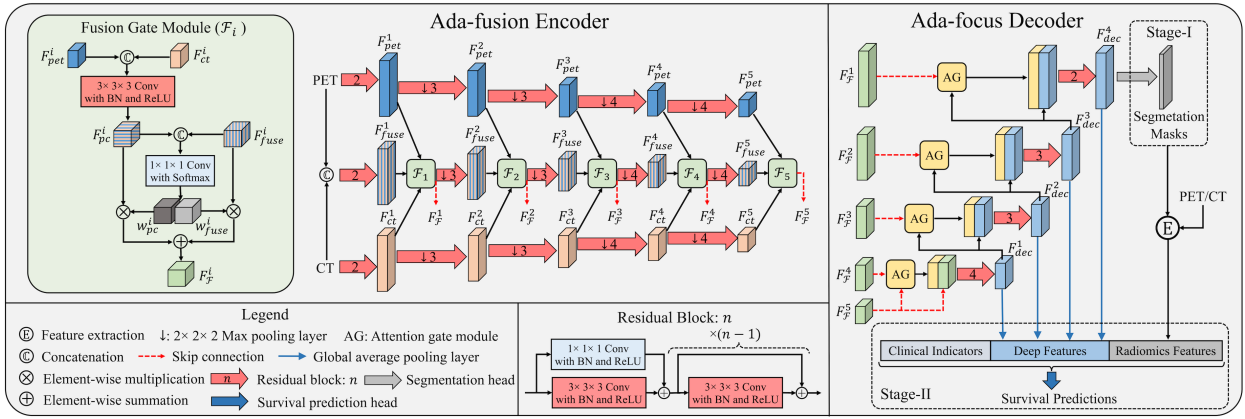

Fig. 2. The architecture of our AdaMSS, which consists of an Ada-fusion encoder and an Ada-focus decoder. The AdaMSS is an adaptive model for tumor segmentation (Stage-I) and survival prediction (Stage-II). The Ada-fusion encoder adopts our data-driven fusion strategy that is adaptively optimized during training, while the Ada-focus decoder adopts attention gate modules to adaptively focus on task-related target regions.

To realize a data-driven fusion strategy, we propose a fusion gate module to control how to fuse $F_{pet}^i$, $F_{ct}^i$, and $F_{fuse}^i$. For the fusion gate module $\mathcal{F}_i$, the $F_{pet}^i$ and $F_{ct}^i$ are first fused to be $F_{pc}^i$ by a 3×3×3 convolutional layer. Then, the $F_{pc}^i$ and $F_{fuse}^i$ are concatenated and mapped into two adaptive weight maps $w_{pc}^i$ and $w_{fuse}^i$ by two 1×1×1 convolutional layers and softmax function. With $w_{pc}^i$ and $w_{fuse}^i$, we can derive $F_{\mathcal{F}}^i = w_{pc}^i F_{pc}^i + w_{fuse}^i F_{fuse}^i$ as the output of $\mathcal{F}_i$. Through fusion gate modules, our AdaMSS can adaptively optimize its fusion strategy during training. All fusion gate modules serve as gates to control how to fuse information, where early, intermediate, and late fusion can be potentially employed as needed. For example, if all the learned $w_{fuse}^i$ are one-maps, the $F_{\mathcal{F}}^i$ will be fully determined by $F_{fuse}^i$ and early fusion will be employed; if all the learned $w_{fuse}^i$ are zero-maps, late fusion will be employed. However, as the $w_{pc}^i$ and $w_{fuse}^i$ are not explicitly encouraged to be zero-/one-maps, the AdaMSS usually employs all early, intermediate, and late fusion but with different weights. We expect our AdaMSS can search over a large possible space of fusion strategies and ultimately find the optimal fusion strategy during training, where the found fusion strategy could vary depending on different training stages (i.e., Stage-I or Stage-II) or different datasets.

### 3.3. Adaptive Focus (Ada-focus) Decoder

The Ada-focus decoder integrates the features from the Ada-fusion encoder and gradually screens out the task-related features via attention gate modules (Schlemper et al., 2019). Specifically, the Ada-focus decoder has a branch $B_{Dec}$ shared by both tumor segmentation and survival prediction. The $B_{Dec}$ is symmetric to the encoder branches $B_{PET/CT}$ and is also composed of successive residual blocks. We embedded an attention gate module before each residual block, which filters the features $F_{\mathcal{F}}^i$ propagated through skip connections. By use of contextual information extracted from their former residual blocks, attention gate modules can spatially depress the non-relevant regions and highlight the task-related regions. The features from the attention gate modules are concatenated with upsampled features derived from former residual blocks and then are fed into the next residual block for further feature screening. Here, we let $F_{dec}^i$ denote the features from the $i^{th}$ residual block of $B_{Dec}$.

In Stage-I, the features from the last residual block of $B_{Dec}$ are fed into a segmentation head that predicts segmentation masks by a 1×1×1 convolutional layer and sigmoid activation. The predicted masks can be used to extract handcrafted radiomics features, which can be leveraged in Stage-II as supplementary information of tumor characteristics. We followed Meng et al. (2022a) to extract 1,456 radiomics features from PET/CT images via Pyradiomics (Van Griethuysen et al., 2017) and selected the most discriminative features through LASSO regression (Tibshirani, 1998). The process of radiomics feature extraction is detailed in Appendix A.

In Stage-II, the features from all residual blocks of $B_{Dec}$ are fed into Global Averaging Pooling (GAP) layers, resulting in a set of deep features. These deep features can be further concatenated with the discriminative radiomics features (derived from Stage-I) and prognostic clinical indicators (e.g., age, gender, TNM stage) for integration. The resultant features, integrating both PET/CT and clinical information, are fed into a survival prediction head, which maps these features into $N$-dimensional vectors using two fully-connected layers with dropout, L2 regularization, and sigmoid activation. The output vectors are the conditional probabilities of patients surviving in $N$ time intervals.

### 3.4. Segmentation-to-Survival Learning

The AdaMSS is first pretrained for tumor segmentation (in Stage-I) and then fine-tuned for survival prediction (in Stage-II). For tumor segmentation, the loss $\mathcal{L}_{Seg}$ is the sum of Dice loss (Milletari et al., 2016) and Focal loss (Lin et al., 2017) between each segmentation prediction and its corresponding ground-truth label. For survival prediction, the loss $\mathcal{L}_{Surv}$ is a negative log-likelihood function (Gensheimer and Narasimhan, 2019):

$$\mathcal{L}_{Surv} = -\sum_{i=1}^{N} \log\left(\max\left(1 + S_i\left(S_i^{pred} - 1\right), \varepsilon\right)\right) + \log\left(\max\left(1 - \bar{S}_i S_i^{pred}, \varepsilon\right)\right), \tag{1}$$

where $S^{pred}, S, \bar{S} \in \mathbb{R}^N$. The $N$ is the number of time intervals and was empirically set as 10 in this study. The time intervals are manually given to ensure that the survival time of all training samples can be evenly distributed in each interval. The $S^{pred}$ is the output of the survival prediction head, while the $S$ and $\bar{S}$ are two label vectors generated from the ground-truth survival outcomes (time-to-event time and censored-or-not status). For $S$, all the time intervals preceding the events are set to 1 while others are 0; For $\bar{S}$, the time interval with the event occurring (only for uncensored patients) is set to 1 while others are 0. With the predicted $S^{pred}$, the estimated survival time can be calculated with:

$$Time^{pred} = \sum_{k=1}^{N}(\prod_{i=1}^{k} S_i^{pred}) \times T_k, \quad (2)$$

where $T \in \mathbb{R}^N$ is the duration of $N$ time intervals.

## 4. Experimental Setup

### 4.1. Datasets and Preprocessing

The OPC dataset consists of 488 histologically proven oropharyngeal H&N cancer patients, which were acquired from the training dataset of the HECKTOR 2022 challenge (Andrearczyk et al., 2023) while the testing dataset was excluded as its ground-truth labels were not released to the public. All patients underwent pretreatment PET/CT and have clinical indicators including age, gender, weight, tobacco/alcohol consumption, Human Papilloma Virus (HPV) status, performance status (Zubrod), and therapy regimens (radiotherapy, chemotherapy, and/or surgery). RFS labels (time-to-event in days and censored-or-not status) were provided as ground truth for survival prediction, while manual annotations of primary tumors and metastatic lymph nodes were provided as ground truth for segmentation. These patients were acquired from seven medical centers (CHUP, CHUS, MDA, HGJ, HMR, CHUM, and CHUV). The patients from two medical centers (CHUM and CHUV) were used for testing and the patients from the remaining five medical centers were used for training, which split the data into 386/102 patients as training/testing sets.

The NPC dataset contains 892 histologically proven NPC patients acquired from Fudan University Shanghai Cancer Center (FUSCC) and Shanghai Proton and Heavy Ion Center (SPHIC). FUSCC and SPHIC Ethical Committee approved this study with informed consent obtained from all included patients. All patients underwent pretreatment PET/CT and have clinical indicators including age, gender, Epstein–Barr Virus (EBV) antibody status, Body Mass Index (BMI), Lactate Dehydrogenase (LDH), histology type, and TNM stage. PFS labels (time-to-event in months and censored-or-not status) were provided as ground truth for survival prediction, while manual annotations of primary tumors were provided as ground truth for segmentation. The patients from FUSCC and SPHIC were used for training and testing respectively, which split the data into 657/235 patients as training/testing sets.

For both OPC and NPC datasets, models were trained and validated using 5-fold cross-validation within the training set and evaluated in the testing set. We performed univariate and multivariate Cox analyses on the clinical indicators so as to screen out the prognostic indicators with significant relevance to RFS/PFS ($P$ <0.05) and leverage them for survival prediction. The distributions of all clinical indicators/characteristics in the OPC and NPC datasets are presented in Appendix B. PET/CT images were resampled into isotropic voxels, where 1 voxel corresponds to 1 mm3, and then cropped to 160×160×160 voxels with the tumor located in the center. PET images were individually standardized using Z-score normalization, while CT images were first clipped to range [−1024, 1024] and then mapped to range [−1, 1].

### 4.2. Implementation Details

Our AdaMSS was implemented using PyTorch on a 12GB NVIDIA Titan V GPU. The Adam optimizer with a batch size of 2 was used for both pretraining (Stage-I) and fine-tuning (Stage-II). The AdaMSS was pretrained for 10,000 iterations in Stage-I and then fine-tuned for another 10,000 iterations in Stage-II. Each training batch included a censored sample and an uncensored sample. In Stage-I, the learning rate was set as 1e-4 initially and then reset to 5e−5 and 1e−5 at the 4,000th and 8,000th training iteration. In Stage-II, the learning rate was set as 5e-5 initially and then reset to 1e−5 and 1e−6 at the 4,000th and 8,000th training iteration. We performed data augmentation in real-time during training to minimize overfitting, including random affine transformations and random cropping to 112×112×112 voxels. Validation was performed after every 200 training iterations and the models achieving the highest validation results in Stage-I and Stage-II were preserved. The model preserved in Stage-I was used to extract radiomics features and initialize the AdaMSS's weights for fine-tuning in Stage-II, while the model preserved in Stage-II was used for the final survival prediction. Our code is publicly available at https://github.com/MungoMeng/Survival-AdaMSS.

### 4.3. Comparison Methods

We compared our AdaMSS with seven existing survival prediction methods, including two traditional radiomics-based methods, two single-task deep survival models, and three multi-task deep survival models. The included traditional methods are Lasso-CoxPH (Tibshirani, 1998) and ICARE (Rebaud et al., 2023). For the traditional methods, radiomics features were extracted from the ground-truth tumor regions using the same feature extraction procedure as Meng et al. (2022a), which is detailed in Appendix A. The included single-task deep survival models are MDSN (Jing et al., 2020) and DeepMTLR-CoxPH (Saeed et al., 2022b). The included multi-task deep survival models are TMSS (Saeed et al., 2022a), DeepMTS (Meng et al., 2022a), and Radio-DeepMTS (Meng et al., 2023). Note that the DeepMTLR-CoxPH, ICARE, and Radio-DeepMTS achieved top performance in the well-benchmarked HECKTOR 2021/2022 challenges, which serve as the most recent state-of-the-art survival prediction methods.

### 4.4. Experimental Settings

The AdaMSS and all comparison methods took the same preprocessed images and clinical indicators as input to ensure a fair comparison. The Concordance index (C-index) and Dice Similarity Coefficient (DSC) were used to evaluate survival prediction and tumor segmentation, which are the commonly used evaluation metrics in related studies (Jing et al., 2020; Andrearczyk et al., 2021; Lu et al., 2021; Meng et al., 2022a; Saeed et al., 2022a; Saeed et al., 2022b; Salahuddin et al., 2023) and also are the standard evaluation metrics for challenges (Andrearczyk et al., 2022; Eisenmann et al. 2022; Andrearczyk et al., 2023). For statistical analysis, a two-sided $P$ value <0.05 is considered to indicate a statistically significant difference.

We also performed two ablation studies to analyze the individual contributions of the proposed SSL strategy and data-driven fusion strategy. Radiomics features were not used in the ablation studies. The proposed learning strategy was compared to the existing Single-Task Learning (STL) and Multi-Task Learning (MTL) strategies, where the AdaMSS trained with different learning strategies were evaluated and compared. The proposed fusion strategy was compared to the existing early, late, and intermediate fusion strategies. For early fusion, the $B_{PET}$, $B_{CT}$, and fusion gate modules were removed; for late fusion, the $B_{Fuse}$ was removed and the $F_{pet}^i$ and $F_{ct}^i$ were propagated to the decoder through skip connections; for intermediate fusion, the fusion gate modules were removed and the $F_{pc}^i$ and $F_{fuse}^i$ were summed.

## 5. Results
### 5.1. Comparison with the State-of-the-art

Table I presents the comparison between our AdaMSS and existing survival prediction methods. For traditional methods, the ICARE achieved higher C-index (0.765/0.689) than the Lasso-CoxPH (0.745/0.671). The single-task deep survival models also achieved higher C-index (MDSN: 0.746/ 0.676 and DeepMTLR-CoxPH: 0.748/0.684) than the Lasso-CoxPH but did not outperform the ICARE. The multi-task deep survival models achieved higher C-index (TMSS: 0.761/ 0.689 and DeepMTS: 0.757/0.695) than the single-task deep survival models and comparable C-index with the ICARE. With the incorporation of radiomics features, the Radio-DeepMTS achieved higher C-index (0.776/0.717) than the DeepMTS and other comparison methods. Nevertheless, our AdaMSS achieved higher C-index (0.804/0.757) than the Radio-DeepMTS and achieved the highest C-index among all methods. Note that significant differences were identified between the AdaMSS and all comparison methods ($P$<0.05).

Our AdaMSS was also compared to its variants that did not integrate radiomics features and/or clinical indicators. Compared with the original AdaMSS, removing radiomics features and/or clinical indicators all resulted in significantly lower C-index (No-radio: 0.780/0.728, No-clinic: 0.795/ 0.749, and No-both: 0.772/0.722; $P$<0.05). Moreover, the segmentation results of the multi-task deep survival models are also reported in Table I. Our AdaMSS also achieved significantly higher DSC (0.775/0.796; $P$<0.05) than the TMSS and DeepMTS (0.754/0.779 and 0.735/0.765). The Radio-DeepMTS had the same segmentation results as the DeepMTS, because the radiomics features were integrated with DeepMTS at the decision level (adopting a CoxPH model to integrate radiomics features and DeepMTS predictions) and were not engaged in the training and inference of DeepMTS.

Table I
Comparison between our AdaMSS and exiting survival prediction methods

| Method | OPC | | NPC | |
|---|---|---|---|---|
| | C-index | DSC | C-index | DSC |
| Lasso-CoxPH | 0.745 * | / | 0.671 * | / |
| ICARE | 0.765 * | / | 0.688 * | / |
| MDSN | 0.746 * | / | 0.676 * | / |
| DeepMTLR-CoxPH | 0.748 * | / | 0.684 * | / |
| TMSS | 0.761 * | 0.754 * | 0.689 * | 0.779 * |
| DeepMTS | 0.757 * | 0.735 * | 0.695 * | 0.765 * |
| Radio-DeepMTS | 0.776 * | 0.735 * | 0.717 * | 0.765 * |
| AdaMSS (Ours) | **0.804** | **0.775** | **0.757** | **0.796** |
| AdaMSS (No-radio) | 0.780 * | **0.775** | 0.728 * | **0.796** |
| AdaMSS (No-clinic) | 0.795 * | **0.775** | 0.749 * | **0.796** |
| AdaMSS (No-both) | 0.772 * | **0.775** | 0.722 * | **0.796** |

**Bold**: the highest C-index and DSC results in each dataset are in bold.
No-radio/clinic/both: without using radiomics features, clinical indicators or both.
*: $P$<0.05, in comparison to AdaMSS (Ours).

### 5.2. Learning Strategy Analysis

Table II shows the results of AdaMSS using different learning strategies for training. The AdaMSS using the STL strategy (segmentation only and survival only) achieved DSC of 0.775/ 0.796 and C-index of 0.754/0.692 on the OPC/NPC datasets, which are regarded as the baseline. Compared with the baseline, the MTL strategy enabled AdaMSS to achieve higher C-index (0.765/0.706) and lower DSC (0.770/0.789). By using our SSL strategy, the AdaMSS achieved significantly higher C-index (0.780/ 0.728; $P$<0.05) than the results of using other learning strategies. Unlike the MTL strategy, our SSL strategy did not degrade the tumor segmentation performance and achieved the same DSC (0.775/0.796) as the baseline.

Table II
Comparison among different learning strategies

| Learning strategy | | OPC | | NPC | |
|---|---|---|---|---|---|
| | | C-index | DSC | C-index | DSC |
| STL | segmentation only | / | **0.775** | / | **0.796** |
| | survival only | 0.754 * | / | 0.692 * | / |
| MTL | | 0.765 * | 0.770 | 0.706 * | 0.789 |
| SSL (Ours) | | **0.780** | **0.775** | **0.728** | **0.796** |

**Bold**: the highest C-index and DSC results in each dataset are in bold.
STL: single-task learning; MTL: multi-task learning; SSL: segmentation-to-survival learning.
*: $P$<0.05, in comparison to SSL.

## 5.3. Fusion Strategy Analysis

Table III shows the results of AdaMSS using different fusion strategies to combine PET/CT information. The results of using PET or CT alone are also included in Table III as the baseline for comparison. Using PET alone resulted in higher C-index (0.768/0.707) and DSC (0.723/0.756) than CT alone (C-index: 0.644/0.574; DSC: 0.645/0.688). Combining PET and CT did not always contribute to better survival prediction performance: Compared to using PET alone, the early and late fusion strategies resulted in lower C-index (0.758/0.697 and 0.765/0.705), while the intermediate fusion strategy resulted in higher C-index (0.770/0.713). Our data-driven fusion strategy led to significantly higher C-index (0.780/0.728; $P<0.05$) than all other fusion strategies. For tumor segmentation, combining PET and CT consistently contributed to better performance: the early, late, and intermediate fusion strategies all resulted in higher DSC (0.750/ 0.772, 0.759/0.780, and 0.764/0.786) than using PET alone. Nevertheless, our data-driven fusion strategy also led to significantly higher DSC (0.775/0.728; $P<0.05$) than all other fusion strategies.

Table III
Comparison among different fusion strategies

| Fusion strategy | OPC | | NPC | |
| --- | --- | --- | --- | --- |
| | C-index | DSC | C-index | DSC |
| Only PET | 0.768 * | 0.723 * | 0.707 * | 0.756 * |
| Only CT | 0.644 * | 0.645 * | 0.574 * | 0.688 * |
| Early fusion | 0.758 * | 0.750 * | 0.697 * | 0.772 * |
| Late fusion | 0.765 * | 0.759 * | 0.705 * | 0.780 * |
| Intermediate fusion | 0.770 * | 0.764 * | 0.713 * | 0.786 * |
| Data-driven fusion (Ours) | **0.780** | **0.775** | **0.728** | **0.796** |

**Bold**: the highest C-index and DSC results in each dataset are in bold.
*: $P<0.05$, in comparison to data-driven fusion.

## 6. Discussion

Our main findings are: (i) Our AdaMSS outperformed the state-of-the-art survival prediction methods across both OPC and NPC datasets, (ii) Our SSL strategy improved survival prediction over the existing STL and MTL strategies, (iii) Our data-driven fusion strategy improved both survival prediction and tumor segmentation over the commonly adopted empirically-designed fusion strategies, and (iv) Our AdaMSS showed adaptability to different tasks in different training stages by self-adapting its focus regions and fusion strategy during training.

In the comparison among our AdaMSS and existing survival prediction methods (Table I), our AdaMSS outperformed all the comparison methods in both OPC and NPC datasets, which suggests that our AdaMSS has achieved state-of-the-art survival prediction performance. The most widely adopted Lasso-CoxPH achieved the poorest performance and was outperformed by the recent deep survival models. This poor performance is likely attributed to the fact that radiomics features have difficulties in representing the prognostic information outside the segmented tumor regions. However, this does not mean radiomics features are unhelpful. The recent radiomics-based method, ICARE, outperformed the single-task deep survival models (MDSN and DeepMTLR-CoxPH) and performed comparably with the state-of-the-art multi-task deep survival models (TMSS and DeepMTS), which suggests that radiomics features are still promising for the capabilities to represent intra-tumor information. Moreover, through integrating the DeepMTS and radiomics features, the Radio-DeepMTS outperformed all other comparison methods, which further demonstrated the potential to leverage radiomics features to improve survival prediction.

We also evaluated the contributions of radiomics features and clinical indicators to our AdaMSS (Table I) and found that adding radiomics features and/or clinical indicators all contributed to statistically significant improvements in survival prediction, which is consistent with the finding in our previous study (Meng et al., 2023). However, compared to the improvements of Radio-DeepMTS over DeepMTS, radiomics features caused larger improvements in our AdaMSS. There are two possible explanations for the larger improvements. Firstly, our AdaMSS achieved better segmentation performance than the DeepMTS, enabling radiomics features to represent intra-tumor information more precisely. Secondly, our AdaMSS integrated radiomics features through cooperative training and inference, while the Radio-DeepMTS integrated radiomics features merely at the decision level. Through the proposed SSL strategy, our AdaMSS decouples tumor segmentation and survival prediction to different training stages, which provides the feasibility to incorporate radiomics features (from Stage-I) into joint training with survival prediction (in Stage-II), and thus achieving closer cooperation between radiomics and deep learning. In addition, even without using radiomics features, our AdaMSS still achieved better survival prediction and tumor segmentation performance than all the comparison methods, which implies that the AdaMSS itself has superior learning capability. We attribute this superiority to the use of the proposed SSL strategy and data-driven fusion strategy, which can be evidenced by ablation studies.

In the learning strategy analysis (Table II), compared to using the STL strategy, the MTL strategy improved survival prediction but degraded tumor segmentation, which indicates the existence of potential destructive interference. Our SSL strategy overcame this destructive interference by decoupling tumor segmentation and survival prediction to different training stages, thereby surpassing the MTL strategy in both survival prediction and tumor segmentation. Our SSL strategy is based on transfer learning (Zhuang et al., 2020), a widely used learning strategy to leverage shared knowledge between tasks. There have been studies comparing MTL and transfer learning (Weller et al., 2022), but no study explored this for survival prediction and tumor segmentation. In addition, despite the simplicity of the idea of transfer learning, many efforts have been made to enable it to work. For segmentation-to-survival learning, a unified model applicable to two distinct tasks should be designed. To achieve this, we made two key designs in the AdaMSS, rendering it as a model that can adaptively fit into two-stage segmentation-to-survival learning.

The Ada-focus decoder serves as a key design for the AdaMSS to enable adaptive adjustment of its focus regions during training. Fig. 3 visualizes the attention maps produced in the Ada-focus decoder, which shows that our AdaMSS can initially focus on tumor regions (in Stage-I) and then expand its focus to include the prognosis-related regions around the tumors (in Stage-II). However, without being pretrained for tumor segmentation, the AdaMSS using STL for survival prediction cannot precisely focus on tumor regions and is interfered with non-relevant backgrounds. MTL enabled the AdaMSS to focus on tumor regions but also hindered it from focusing on the prognosis-related regions outside tumors. The differences in attention maps (Fig. 3) can partly explain the superior performance of our AdaMSS: it can capture more prognostic information within and around the tumors while getting less interference from non-relevant backgrounds. The captured information around tumors may represent the heterogeneity in tumor periphery (Erber et al., 2022) or the potential adjacent tissue invasion (Yamaguchi et al., 2008; Hao et al., 2018).

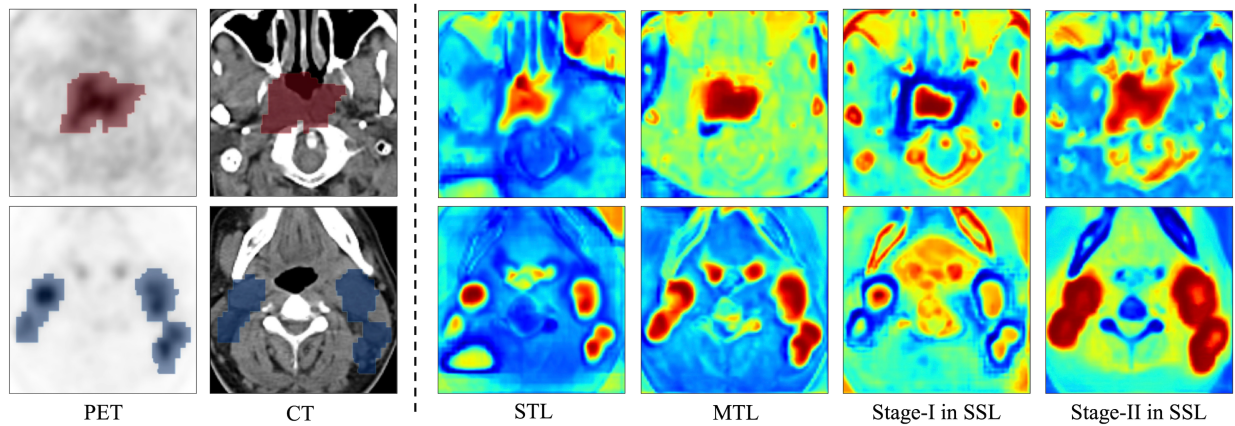

Fig. 3. Visualizations of the attention maps in the Ada-focus decoder. Two cross-sectional slices containing tumor regions (primary tumors and metastatic lymph nodes) are presented in two rows. From left to right are the input PET/CT images and the attention maps generated by the AdaMSS using single-task learning (STL), multi-task learning (MTL), and segmentation-to-survival learning (Stage-I and Stage-II in SSL). The PET/CT images are delineated with primary tumors (red) or metastatic lymph nodes (blue) ground-truth annotations.

The Ada-fusion encoder (data-driven fusion strategy) is another key design to facilitate segmentation-to-survival learning. In the fusion strategy analysis (Table III), the empirically-designed fusion strategies showed inconsistent performance in tumor segmentation and survival prediction, which suggests that these two tasks have different requirements for fusion strategies and the optimal fusion strategy also should vary depending on the tasks. With the data-driven strategy, our AdaMSS can adaptively optimize its fusion strategy during training, which made it achieve the best performance in both survival prediction and tumor segmentation. For survival prediction, the early and late fusion strategies failed to outperform using PET alone, which is consistent with the findings in Wang et al. (2022)'s study. As we have mentioned, this is because early and late fusion have difficulties in extracting intra- and inter-modality information. The intermediate fusion strategy allowed information to be fused at multiple scales and thus achieved better performance. Nevertheless, the intermediate fusion strategy is still limited to manually defined prior knowledge and was defeated by our data-driven fusion strategy. In Table IV, we attempt to interpret the data-driven fusion strategy by calculating the mean values of the adaptive weight maps in fusion gate modules. These mean values are 0.5 initially and then are optimized during training, which could be regarded as indicators representing the ratio of information that each fusion gate module fuses from different sources (i.e., $B_{PET/CT}$ or $B_{Fuse}$). For example, a mean value of $w_{fuse}^i$ lower than 0.5 indicates that the fusion gate module $\mathcal{F}_i$ tends to fuse less information (<50%) from the $F_{fuse}^i$ in $B_{Fuse}$. Through this interpretation, we found that our AdaMSS adapted its fusion strategies for different training stages and datasets, which demonstrates the adaptability and generalizability of our data-driven fusion strategy to different tasks and data.

Our study has a few limitations. First, by using a data-driven fusion strategy, we expect that our AdaMSS can search over a large search space to find the optimal fusion strategy based on training data. However, despite the current search space being vast and including commonly-adopted fusion strategies (early, late, and intermediate fusion), we will investigate further inclusion of other fusion operations, such as cross-modal transformers (Sun et al., 2022), into the search space. Moreover, this study mainly focused on using PET/CT for survival prediction in OPC and NPC. Our AdaMSS can be further validated with other diseases (e.g., brain gliomas, lung cancer) or other multi-modality images (e.g., multi-parametric MRI, PET/MRI).

Table IV
The mean values of the adaptive weight maps in the fusion gate modules of the Ada-fusion encoder.

| $w_{fuse}^i$ with $i =$ | | 1 | 2 | 3 | 4 | 5 |
|---|---|---|---|---|---|---|
| OPC | Stage-I | 0.494 | 0.471 | 0.445 | 0.384 | 0.300 |
|  | Stage-II | 0.525 | 0.487 | 0.429 | 0.308 | 0.398 |
| NPC | Stage-I | 0.490 | 0.463 | 0.427 | 0.362 | 0.274 |
|  | Stage-II | 0.513 | 0.475 | 0.410 | 0.294 | 0.351 |

Note: The mean value of $w_{pc}^i$ is equal to (1 - the mean value of $w_{fuse}^i$).

## 7. Conclusion

In this study, we have presented an Adaptive Multi-modality Segmentation-to-Survival model (AdaMSS) for survival prediction from PET/CT images. The AdaMSS introduces a Segmentation-to-Survival Learning (SSL) strategy for model training and a data-driven strategy for multi-modality image information fusion, which have shown superiority over the existing single- and multi-task learning strategies and the existing empirically-designed fusion strategies. With the SSL and data-driven fusion strategies, our AdaMSS is endowed with the adaptability to optimize its focus regions and fusion strategy during training, thus enabling it to achieve consistent improvements in both survival prediction and tumor segmentation. The experimental results with two large multi-center clinical datasets demonstrated that our AdaMSS can outperform the state-of-the-art survival prediction methods and show generalizability to different diseases (OPC and NPC).

**Appendix A: Radiomics Feature Extraction**

With the segmentation masks predicted in Stage-I, we extracted handcrafted radiomics features from PET/CT images via Pyradiomics (Van Griethuysen et al., 2017). The extracted radiomics features include 19 features from First Order Statistics (FOS), 24 features from Grey-Level Cooccurrence Matrix (GLCM), 16 features from Grey-Level Run Length Matrix (GLRLM), 16 features from Grey-Level Size Zone Matrix (GLSZM), 5 features from Neighboring Grey Tone Difference Matrix (NGTDM), and 16 features based on 3D tumor shape. The shape-based features were extracted from the tumor segmentation mask, while other features were extracted from PET and CT images separately. The PET/CT-derived features were also recomputed for different wavelet decomposition of PET/CT. Performing low-pass (L) or high-pass (H) wavelet filter along x, y, or z directions resulted in eight decompositions (LLL, LLH, LHL, LHH, HHH, HLL, HHL, and HLH). Consequently, we derived a total of 1,456 radiomics features. All radiomics features were standardized using Z-score normalization. The redundant features with Spearman's correlation >0.7 were eliminated, followed by LASSO regression (Tibshirani, 1998) for further feature selection.

**Appendix B: Clinical Characteristics**

Table V
Demographic and clinical characteristics of patients in the OPC dataset.

| Characteristics | Training set | Testing set |
|---|---|---|
| Number of Patients | 386 | 102 |
| Age (year), median (range) | 60 (32-85) | 64 (44-90) |
| Weight (kg), median (range) | 81.5 (41-160) | 74 (34-120) |
| Gender, Number (%) | | |
|     Male | 322 (83.4) | 80 (78.4) |
|     Female | 64 (16.6) | 22 (21.6) |
| Alcohol consumption, Number (%) | | |
|     Yes | 95 (24.6) | 0 (0.0) |
|     No | 59 (15.3) | 0 (0.0) |
|     Unknown | 232 (60.1) | 102 (100.0) |
| Tobacco consumption, Number (%) | | |
|     Yes | 85 (22.0) | 0 (0.0) |
|     No | 105 (27.2) | 0 (0.0) |
|     Unknown | 196 (50.8) | 102 (100.0) |
| HPV status, Number (%) | | |
|     Positive | 252 (65.3) | 22 (21.6) |
|     Negative | 41 (10.6) | 2 (2.0) |
|     Unknown | 93 (24.1) | 78 (76.4) |
| Performance status (Zubrod), Number (%) | | |
|     0 | 86 (22.3) | 0 (0.0) |
|     1 | 114 (29.5) | 0 (0.0) |
|     2 | 11 (2.8) | 0 (0.0) |
|     3 | 3 (0.8) | 0 (0.0) |
|     4 | 1 (0.3) | 0 (0.0) |
|     Unknown | 171 (44.3) | 102 (100.0) |
| Surgery, Number (%) | | |
|     Yes | 50 (13.0) | 0 (0.0) |
|     No | 202 (52.3) | 46 (45.1) |
|     Unknown | 134 (34.7) | 56 (54.9) |
| Chemotherapy, Number (%) | | |
|     Yes | 324 (83.9) | 98 (96.1) |
|     No | 62 (16.1) | 4 (3.9) |
| RFS, Number (%) | | |
|     Uncensored (recurrence) | 81 (21.0) | 15 (14.7) |
|     Censored (recurrence free) | 305 (79.0) | 87 (85.3) |

Table VI
Demographic and clinical characteristics of patients in the NPC dataset.

| Characteristics | Training set | Testing set |
|---|---|---|
| Number of Patients | 657 | 235 |
| Age (year), median (range) | 46 (14-83) | 48 (14-74) |
| BMI (Kg/m2), mean (range) | 23.26 (14.69-38.89) | 24.09 (16.41-34.38) |
| LDH (U/L), mean (range) | 193.96 (89-782) | 212.89 (111-1400) |
| Gender, Number (%) | | |
| Male | 508 (77.32) | 180 (76.60) |
| Female | 149 (22.68) | 55 (23.40) |
| EBV antibody, Number (%) | | |
| Positive | 418 (63.62) | 155 (65.96) |
| Negative | 125 (19.03) | 42 (17.87) |
| Unknown | 114 (17.35) | 38 (16.17) |
| Histology (WHO Type), Number (%) | | |
| I | 6 (0.91) | 2 (0.85) |
| II | 64 (9.74) | 6 (2.55) |
| III | 587 (89.35) | 227 (96.60) |
| T stage, Number (%) | | |
| T1 | 168 (25.57) | 68 (28.94) |
| T2 | 70 (10.65) | 29 (12.34) |
| T3 | 357 (54.34) | 129 (54.89) |
| T4 | 62 (9.44) | 9 (3.83) |
| N stage, Number (%) | | |
| N0 | 30 (4.57) | 6 (2.55) |
| N1 | 165 (25.11) | 54 (22.98) |
| N2 | 379 (57.69) | 138 (58.72) |
| N3 | 83 (12.63) | 37 (15.75) |
| TNM stage, Number (%) | | |
| III | 518 (78.84) | 190 (80.86) |
| IVa | 139 (21.16) | 45 (19.15) |
| PFS, Number (%) | | |
| Uncensored (progression) | 151 (22.98) | 45 (19.15) |
| Censored (progression free) | 506 (77.02) | 190 (80.85) |

Note: WHO Type I = keratinizing; II = non-keratinizing (differentiated); III = non-keratinizing (undifferentiated).

## Acknowledgements

This work was supported by Australian Research Council (ARC) under Grant DP200103748.## References

Andrearczyk, V., Fontaine, P., Oreiller, V., Castelli, J., Jreige, M., Prior, J.O., Depeursinge, A., 2021, Multi-task deep segmentation and radiomics for automatic prognosis in head and neck cancer, in: Rekik, I., et al. (Eds.), Predictive Intelligence in Medicine, pp. 147–156. https://doi.org/10.1007/978-3-030-87602-9_14.

Andrearczyk, V., Oreiller, V., Boughdad, S., Rest, C.C.L., Elhalawani, H., Jreige, M., Prior, J.O., Vallières, M., Visvikis, D., Hatt, M., Depeursinge, A., 2022, Overview of the HECKTOR Challenge at MICCAI 2021: Automatic Head and Neck Tumor Segmentation and Outcome Prediction in PET/CT Images, in: Andrearczyk, V., et al. (Eds.), Head and Neck Tumor Segmentation and Outcome Prediction, pp. 1-37. https://doi.org/10.1007/978-3-030-98253-9_1.

Andrearczyk, V., Oreiller, V., Abobakr, M., Akhavanallaf, A., Balermpas, P., Boughdad, S., Capriotti, L., Castelli, J., Le Rest, C.C., Decazes, P., Correia, R., 2023, Overview of the HECKTOR Challenge at MICCAI 2022: Automatic Head and Neck Tumor Segmentation and Outcome Prediction in PET/CT Images, in: Andrearczyk, V., et al. (Eds.), Head and Neck Tumor Segmentation and Outcome Prediction, pp. 1-30. https://doi.org/10.1007/978-3-031-27420-6_1.

Bi, L., Feng, D.D., Fulham, M., Kim, J., 2020. Multi-label classification of multi-modality skin lesion via hyper-connected convolutional neural network. Pattern Recognition. 107, 107502. https://doi.org/10.1016/j.patcog.2020.107502.

Cox, D.R., 1972. Regression models and life-tables. Journal of the Royal Statistical Society: Series B (Methodological). 34(2), 187-202. https://doi.org/10.1111/j.2517-6161.1972.tb00899.x.

Cui, C., Cui, C., Liu, H., Liu, Q., Deng, R., Asad, Z., Wang, Y., Zhao, S., Yang, H., Landman, B.A., Huo, Y., 2022, Survival Prediction of Brain Cancer with Incomplete Radiology, Pathology, Genomic, and Demographic Data, in: Wang, L., et al. (Eds.), Medical Image Computing and Computer Assisted Intervention – MICCAI 2022, pp. 626-635. https://doi.org/10.1007/978-3-031-16443-9_60.14